\documentclass[usenatbib]{mn2e}
\usepackage{graphicx}
\usepackage{longtable}
\input psfig.sty

\title[The Evolution and Star Formation History of M33]
    {The Evolution and Star Formation History of M33}

\author[X. Y. Kang et al. ]
{Xiaoyu~Kang$^{1,2,3,4}$\thanks{E-mail: kxyysl@ynao.ac.cn},
 Ruixiang~Chang$^{3}$, Jun~Yin$^3$, Jinliang~Hou$^3$, Fenghui~Zhang$^{1,2}$, \and Yu~Zhang$^{1,2,4}$ and Zhanwen~Han$^{1,2}$\\
$^1$National Astronomical Observatories, Yunnan Observatory, Chinese Academy of Sciences, Kunming, 650011, China\\
$^2$Key Laboratory for the Structure and Evolution of Celestial Objects, Yunnan Astronomical Observatory, Chinese Academy of Sciences,\\
 Kunming, 650011, China\\
$^3$Key Laboratory for Research in Galaxies and Cosmology, Shanghai Astronomical Observatory, Chinese Academy of Sciences, 80 Nandan \\
Road, Shanghai, 200030, China \\
$^4$Graduate School of the Chinese Academy of Sciences, Beijing 100039, China\\
   }

\begin{document}

\date{\today}

\pagerange{\pageref{firstpage}--\pageref{lastpage}}

\pubyear{2012}

\maketitle

\label{firstpage}

\begin{abstract}

We construct a parameterized model to explore
the main properties of the star formation history of M33. We assume
that the disk originates and grows by the primordial gas infall and
adopt the simple form of gas accretion rate with one free parameter,
the infall time-scale. We also include the contribution of gas
outflow process.  A major update of the model is that
we adopt a molecular hydrogen correlated star formation law and
calculate the evolution of the atomic and molecular gas separately.
Comparisons between the model predictions and the observational data
show that the model predictions are very sensitive to the adopted
infall time-scale, while the gas outflow process mainly influences
the metallicity profile. The model adopting a moderate outflow rate
and an inside-out formation scenario can be in good agreement with
most of observed constraints of M33 disk. We also compare the model
predictions based on the molecular hydrogen correlated star
formation law and that based on the Kennicutt star formation law.
Our results imply that the molecular hydrogen correlated
star formation law should be preferred to describe the evolution
of the M33 disk, especially the radial distributions of both the
cold gas and the stellar population.

\end{abstract}


\section{Introduction}

The NGC 598 (M33) galaxy is a low-luminosity, late-type disk galaxy
in the Local Group. M33 is observed to be much smaller and less
massive than the Milky Way galaxy, but has much larger gas fraction.
It also shows no signs of recent mergers and no presence of prominent bulge
and bar component (Regan \& Vogel 1994; McLean \& Liu 1996). In addition,
due to its proximity, large angular size, and rather low inclination,
M33 is an excellent target for detailed observations of its cold
gas, metallicity, the star formation rate (SFR) and stellar population,
and thus provides an excellent chance for testing the model of
galactic chemical evolution.

The star formation (SF) law is one of the important ingredients of
the model. Based on the observed data of a sample of 97 nearby normal
and star-burst galaxies, Kennicutt (1998) found a power-law correlation between
the galaxy-averaged SFR surface density ($\Psi(r,t)$) and the galaxy-averaged
total gas surface density ($\Sigma_{\rm gas}(r,t)$), which was termed as the classical
Kennicutt SF law. Later, observations of high spatial resolution
(less than kpc-scale regions) showed that the SFR surface density
correlated stronger with the surface density of the molecular hydrogen
($\Sigma_{\rm H_2}(r,t)$) than with that of the atomic hydrogen
($\Sigma_{\rm HI}(r,t)$) and the total gas
(Wong \& Blitz 2002; Kennicutt et al. 2007;
Bigiel et al. 2008; Leroy et al. 2008). It was also shown that the SFR
surface density is almost proportional to $\Sigma_{\rm H_2}(r,t)$:
\begin{equation}
\Psi(r,t)=\Sigma_{\rm{H_2}}(r,t)/t_{\rm dep},
\label{eq:h2sfr}
\end{equation}
where $t_{\rm dep}$ is the molecular hydrogen depletion time.
Hereafter, Equ. \ref{eq:h2sfr} is called as the $\Sigma_{\rm H_2}$-based SF law
in this paper.

Moreover, the gas outflow process may influence the evolution of M33.
Garnett (2002) concluded that spiral
galaxy with $V_{\rm rot}\leq125\rm km\,s^{-1}$ may lose some
amount of gas in supernova-driven winds and Tremonti et al. (2004)
also confirmed this conclusion. The results of Chang et al. (2010)
indicated that the gas outflow process plays an important role in the
chemical evolution of the disk galaxy since it can bring part of
newly formed metal off the galactic disk. They show that the model
assuming that the gas outflow efficiency increases as its stellar
mass decreases can explain the observed mass-metallicity relation.

The chemical evolution of M33 has been studied by several groups in
previous studies (Moll\'{a} et al. 1996; Magrini et al. 2007a; Marcon-Uchida
et al. 2010). Magrini et al. (2007a) found that the model adopting an
almost constant gas-infall rate can reproduce some of the observed
properties of M33, especially the observed relatively high SFR and
the shallow abundance gradients. Marcon-Uchida et al. (2010)
compared the chemical evolution of the Milky Way, M31 and M33.
They found that the model predictions of the Milky Way and
that of M31 were in good agreement with the main features of
observations, while the model of M33 failed to reproduce the
present-day gas surface density in the inner disk. The oxygen
abundance was also overestimated by 0.25 dex in the whole disk of
M33.

In this paper, we build a bridge between the observed data of M33 and
its star formation history (SFH) by constructing a parameterized model
of its formation and evolution. A major update of the model is that
we adopt the $\Sigma_{\rm H_2}$-based SF law and this is maybe
the first time for the model of M33 to  calculate separately
the evolution of the atomic and molecular gas.

The paper is organized as follows. Section 2 describes the observed
features of M33 disk, including the surface brightness, the SFR,
the cold gas content and the metallicity etc.. The main assumptions and
ingredients of our model are presented in Section 3. The comparisons
between the model predictions and the observations are shown in Section 4,
and Section 5 summarizes our main conclusions.

\section{The Observations}
\label{sect:Obs}

In this section, we summarize the current available
observations of M33 galaxy, especially the radial
distributions along the disk, including the surface densities of gas
and SFR, the surface brightness, the color, and the metallicity. Our
model predictions will be compared with all these observed
trends.

\subsection{Surface brightness and radial distribution of colors}

\begin{figure}
  \centering
  \includegraphics[angle=0,height=8cm,width=8cm]{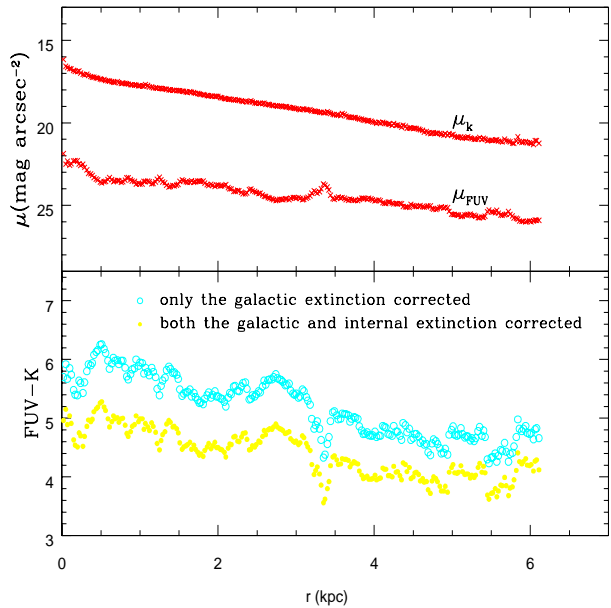}\\
    \caption{Surface brightness and color radial profiles of M33 disk
    (Mu\~{n}oz-Mateos et al. 2007).
    \emph{upper panel}: the surface brightness profiles of FUV$-$band and $K-$band.
    \emph{lower panel}: the color profile of FUV$-K$.
    The observed data are described in details in Section 2.1. }
  \label{Fig:color}
\end{figure}

\begin{table}
\noindent Table 1. M33 disk scale-length (distance=840\,kpc). \\
[2mm]
\begin{tabular}{lll}
\hline \hline
  Tracer   & $r_{\rm d}$ (kpc)      &   Refs.     \\ [1mm]
\hline
Gas $^{a}$ &         &             \\

 CO(1-0)   & 2.5         &  1       \\
 CO(1-0)   & 2.0         &  2       \\
 CO(1-0)   & 1.4         &  3       \\
 CO(2-1)   & 1.4         &  4       \\
 CO(2-1)   & 1.9         &  5       \\
 CO+21cm   & 7.8 $^{b}$  &  1, 6     \\
 \hline
  young stellar pop                 \\
  (0.5 $\sim$ 6 kpc)            \\
 NUV       & 2.10        &  7 (Galex)     \\
 FUV       & 2.20        &  7 (Galex)    \\
 H$_\alpha$      & 1.80        &  7, 8       \\
 {\it B} band    & 1.9         &  9           \\
 {\it V} band   & 1.89        &  10 (HST ACS)     \\
\hline
  older stellar pop                 \\
  (0.5 $\sim$7 kpc)   \\
 {\it K} band    & 1.00        &  4 (2MASS)    \\
 {\it K} band    & 1.4         &  9            \\
 3.6$\,\mu$m & 1.56        &  7 (Spitzer IRAC)  \\
 4.5$\,\mu$m & 1.55        &  7 (Spitzer IRAC)   \\
\hline
  dust, diffuse emmision                   \\
  (0.5 $\sim$7 kpc)               \\
 5.8$\,\mu$m & 1.61        &  7 (Spitzer IRAC)   \\
 8.0$\,\mu$m & 1.44        &  4, 7 (Spitzer IRAC) \\
 24$\,\mu$m  & 1.55        &  4 (Spitzer MIPS)   \\
 24$\,\mu$m  & 1.40        &  5 (Spitzer MIPS)   \\
 24$\,\mu$m  & 1.77        &  7 (Spitzer MIPS)   \\
 60$\,\mu$m  & 1.30        &  11 (ISO)   \\
 70$\,\mu$m  & 1.48        &  5 (Spitzer MIPS)    \\
 70$\,\mu$m  & 1.74        &  7 (Spitzer MIPS)    \\
 160$\,\mu$m & 1.83        &  5 (Spitzer MIPS)    \\
 160$\,\mu$m & 1.99        &  7 (Spitzer MIPS)    \\
 170$\,\mu$m & 1.80        &  11 (ISO)  \\
 \hline \hline
\end{tabular} \\ [1mm]
a: assuming $\Sigma_{\rm gas} \propto {\rm exp}(-r/r_{\rm d})$.\\
b: assuming $\Sigma_{\rm gas} \propto {\rm exp}[-(r/r_{\rm d})^2]$. \\
Refs: (1) Corbelli (2003); (2) Heyer et al. (2004); (3) Engariola et al.
(2003); (4) Gardan et al. (2007); (5) Gratier et al. (2010); (6) Magrini
et al. (2007a); (7) Verley et al. (2009); (8) Hoopes \& Walterbos (2000);
(9) Regan \& Vogel (1994); (10) Williams et al. (2009); (11) Hippelein
et al. (2003).

\end{table}

The surface brightness observed in multi-bands of a galaxy
contains important information of its SFH. For example, since
the surface brightness in the FUV$-$band is very sensitive to the
presence of recent SF activity and that in the near-infrared $K-$band
strongly correlates with the accumulated SF in the galaxy, the
observed radial distribution of FUV$-K$ color provides tight
constraints on the specific SFR of the galaxy.

Mu\~{n}oz-Mateos et
al. (2007) derived the M33's azimuthally averaged radial profiles in
FUV$-K$ color by using Galaxy Evolution Explorer (GALEX) UV bands
surface brightness (Gil de Paz et al. 2007) and a deep (Two-Micron
All-Sky Survey) 2MASS image ($K<{\rm21\,mag\,arcsec^{-2}}$).
We show the measured FUV$-$band and $K-$band surface brightness as well
as the FUV$-K$ color profiles from Mu\~{n}oz-Mateos et al. (2007) in
Fig. \ref{Fig:color}. In the upper panel, we plot the surface
brightness of FUV$-$band and $K-$band as the red crosses which have been
corrected only for the Galactic extinction. The lower panel shows
the radial profiles of FUV$-K$ color, the cyan open circles have been
only corrected for the Galactic extinction, and the yellow filled
circles have been corrected for the both Galactic and internal
extinction. It can be seen from Fig. \ref{Fig:color} that the
negative gradient (after the extinction correction) implies an
inside-out formation process, that is, the stellar populations become
relatively blue and young as the radius increases.


The disk scale-length $r_{\rm d}$ can be obtained from radial
profile, but it has some complexity since it is not only dependent
on the wavelength, but also related to what we refer: gas disk, stellar disk
or the total. Based on the surface brightness profiles in $B$ and $K$
bands, Regan \& Vogel (1994) derived a stellar scale-length of
$1.9\rm\,kpc$ and 1.4$\rm\,kpc$ respectively. While the CO
(representing the molecular hydrogen distribution) radial scale-length is
about $2.5\rm\,kpc$ (Corbelli 2003), larger than that of the
stellar disk. When the total gas mass surface density
($\Sigma_{\rm H_2}+\Sigma_{\rm HI}$) is concerned, the
resulted scale-length is much larger (Corbelli 2003; Magrini et al.
2007a). In Table 1, we list the scale-length for different disk
components of the M33. In general, the $K-$band luminosity reflects
the stellar profile, a smaller value of $r_{\rm d,K}$ implies that the
stellar disk is more concentrated than the gas disk.
In this paper we will adopt $r_{\rm d}=1.4\rm\,kpc$ (Regan \& Vogal
1994). The total stellar mass of M33 is estimated to be
$\sim(3.0-6.0)\times10^{9}\,{\rm M}_{\odot}$ (Corbelli 2003).

\subsection{Profiles of cold gas, SFR and gas depletion time}

\begin{figure*}
   \centering
   \includegraphics[angle=0,width=0.75\textwidth]{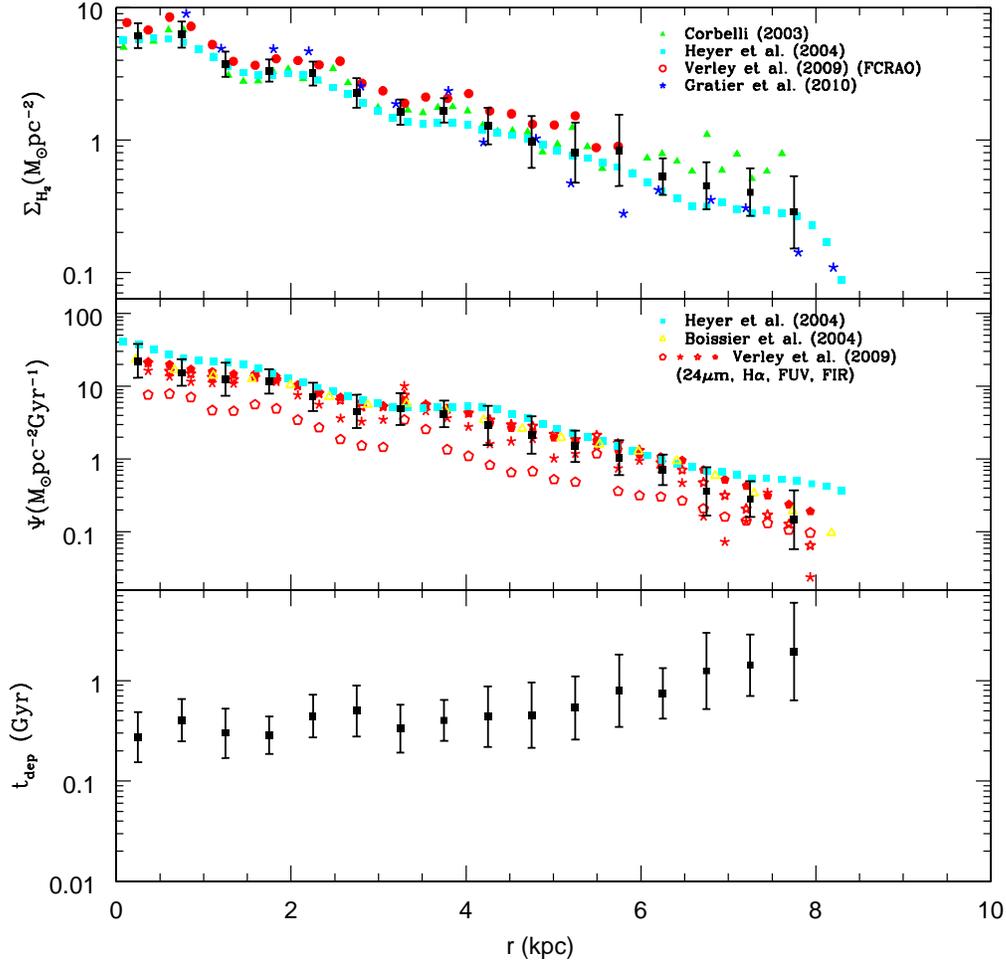}
   \caption{Observed profiles of the surface density of $\rm H_2$
   and SFR, and the molecular gas depletion time for M33 disk.
\emph{top panel}: current surface density profiles of $\rm H_2$.
\emph{middle panel}: current surface density profiles of SFR.
The mean values of both $\Psi(r,t)$ and $\Sigma_{\rm H_2}$
in each bin are shown as filled squares with error bars, where the error bars
represent the distribution dispersion in each bin.
\emph{bottom panel}: the filled squares are the molecular gas depletion time
estimated according to the $\Sigma_{\rm H_2}$-based SF law, and the
error bars also represent the distribution dispersion in each bin.
The observed data are described in details in Section 2.2.}
   \label{Fig:dep}
   \end{figure*}

During the past years, a number of data sets relating to the atomic
and molecular gas distributions in M33 are becoming available.
Imaging of molecular clouds has recently been carried out in M33
with the Berkeley-Illinois-Maryland Association (BIMA)
interferometer and with the Five College Radio Astronomy Observatory
(FCRAO) 14m telescope using the CO $J=1-0$ line transition (Corbelli
2003; Engargiola et al. 2003; Heyer et al. 2004; Verley et al.
2009). It was also observed by the Institute de RadioAstronomie
Millim\'{e}trique (IRAM) 30m telescope using the CO $J=2-1$ line
transition (Gardan et al. 2007; Gratier et al. 2010). The observed
radial profiles of H$_{2}$ surface density are plotted in the top
panel of Fig. \ref{Fig:dep}, where the data are taken from Corbelli
(2003) (the green filled triangles), Heyer et al. (2004) (the cyan
filled squares), Verley et al. (2009) (the red filled circles
(FCRAO)), and Gratier et al. (2010) (the blue asterisks).


The cold gas distribution is tightly correlated with the
distribution of the SFR. Several groups have already measured the
SFR in several regions along the disk of M33 using different
tracers, including the H$\alpha$ emission, the luminosity in the
far-ultraviolet (FUV) band and the far-infrared (FIR) band
(Hippelein et al. 2003; Engargiola et al. 2003; Heyer et al. 2004;
Gardan et al. 2007; Boissier et al. 2007; Verley et al. 2009). The
observed data of the radial distribution of the SFR surface
densities are plotted in the middle panel of Fig. \ref{Fig:dep}. The
data taken from Heyer et al. (2004) and Boissier et al. (2007) are
shown by the cyan filled squares (FIR) and the yellow empty
triangles (FUV), respectively. The data taken from Verley et al.
(2009) are represented by the red empty pentagons (24$\,\rm \mu m$), the
red asterisks (H$\alpha$), the red five-pointed stars (FUV), and the
red filled pentagon (FIR).

Since the observed SFR surface density $\Psi(r,t)$ and the
molecular hydrogen surface density $\Sigma_{\rm H_2}(r,t)$ are both available, it
is easy to estimate the molecular gas depletion time
$t_{\rm dep}$ according to the $\Sigma_{\rm H_2}$-based SF law.
In practice, we first divide the observed
data along the disk of M33 into 16 bins. Then, we calculate the mean values
of both $\Psi(r,t)$ and $\Sigma_{\rm H_2}(r,t)$ in each bin in the
logarithmic coordinates and show them as filled squares with error
bars in top and middle panels of Fig. \ref{Fig:dep},
where the error bars represent the distribution dispersion in each
bin. Using the mean values obtained, we calculate the
molecular gas depletion time in each bin through
$t_{\rm dep}=\Sigma_{\rm H_2}(r,t)/\Psi(r,t)$ (according to the
$\Sigma_{\rm H_2}$-based SF law) and plot them in the
bottom panel of Fig. \ref{Fig:dep} as filled squares with error
bars. It can be seen that $t_{\rm dep}$ is almost constant along
the most part of the disk except in the outer region of the disk
($r>4r_{\rm d}$), where $t_{\rm dep}$ increases gradually as
the galactic radius increases. For the purpose of simplicity,
we adopt the mean value of the molecular gas depletion time
and fix $t_{\rm dep}=0.43\,\rm{Gyr}$ throughout this work.

The reciprocal quantity of the molecular gas depletion time
$1/t_{\rm dep}$ of a galaxy correlates to its star formation
efficiency (SFE), which describes the proportion of molecular
gas turns into stellar mass in unit time. Leroy et al. (2008)
explored the SFEs of 23 nearby galaxies, and they estimated
that the average value of $t_{\rm dep}$ is about
$1.9\,\rm{Gyr}$ in large spirals. However, the case of M33
is slightly different. Gardan et al. (2007) found that M33 was more
efficient in forming stars than large spirals in the local universe
and the molecular depletion time was about $0.11 -
0.32\,\rm{Gyr}$. Gratier et al. (2010) also concluded that the SFE
of M33 appeared to be 2-4 times higher than that observed in massive
spiral galaxies. Our results of Fig. \ref{Fig:dep} also suggest that
the SFE of M33 is almost constant along the disk and higher than the
average value of SFE of large spirals investigated by Leroy et al. (2008).

The H{\sc i} surface density profile has been derived from high
sensitivity observations with the Westerbork Synthesis Radio
Telescope (WSRT)+Effelsberg 100m (Deul \& van der Hulst 1987), with
the Arecibo (Corbelli \& Schneider 1997; Putman et al. 2009) and
with the Very Large Array (VLA) B, C, and D (Gratier et al. 2010).

\subsection{Disk abundance gradients}

\begin{table}
{\bf Table 2}. Observed abundance gradients in M33 (dlog(O/H)/dR, in
$\rm dex\,kpc^{-1}$).
\\ [2mm]
\begin{tabular}{llll}
\hline \hline
Objects  &     $R_{\rm g}$   &     Oxygen gradient        &      Refs.  \\
         &    (kpc)          &     $\rm dex\,kpc^{-1}$    &            \\
\hline
 H{\sc ii}     & 1.0--5.7   &  $-$0.13              &  (1)  \\
         & 0.2--6.5   &  $-$0.070$\pm$0.008   &  (2)  \\
         & 0.4--6.5   &  $-$0.127$\pm$0.011   &  (3)  \\
         & 0.3--11.0  &  $-$0.19 $\pm$0.03    &  (4)  \\
         & 1.0--7.2   &  $-$0.012$\pm$0.011   &  (5)  \\
         & 2.0--7.2   &  $-$0.054$\pm$0.011   &  (6)  \\
         & 0.2--6.0   &  $-$0.027$\pm$0.012   &  (7)   \\
         & 1.0--8.0   &  $-$0.044$\pm$0.009   &  (8)   \\
\hline
 PNe     & 1.0--10.0  &  $-$0.11$\pm$0.04    &  (9)  \\
         & 0.5--8.5   &  $-$0.031$\pm$0.013   &  (10)  \\
         & 0.5--8.0   &  $-$0.013$\pm$0.016   &  (11)  \\
\hline
 B stars & 0.2--4.1   &  $-$0.16$\pm$0.06     &  (12)  \\
         & 0.25--8.0  &  $-$0.06$\pm$0.02     &  (13)  \\
\hline \hline

\end{tabular} \\[2mm]
Refs: (1) Kwitter \& Aller (1981); (2) V\'{\i}lchez et al. (1988); (3)
Zaritsky et al. (1994); (4) Henry \& Howard (1995);
(5) Crockett et al. (2006);
(6) Magrini et al. (2007b); (7) Rosolowsky \& Simon (2008);
(8) Magrini et al. (2010); (9) Magrini et al. (2007a);
(10) Magrini et al. (2009); (11) Bresolin et al. (2010);
(12) Monteverde et al. (1997); (13) Urbaneja et al. (2005).

\end{table}

\begin{figure}
  \centering
  \includegraphics[angle=0,height=6cm,width=8cm]{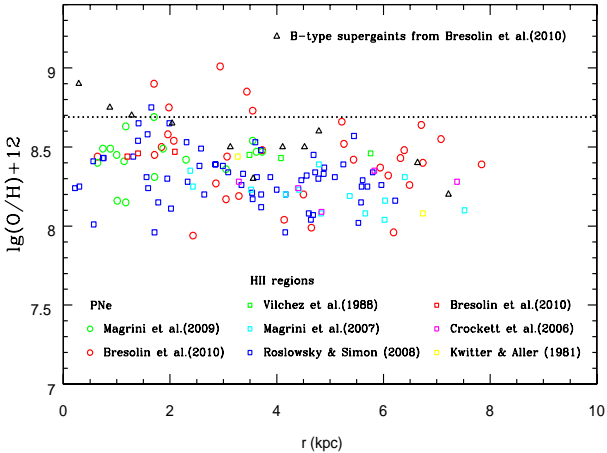}\\
    \caption{Observed O abundance in M33 disk. Different data notations
    represent those taken from various literature. Dotted line in the
    panel is the corresponding solar values taken from Asplund et al.
    (2009). The observed data are described in details in Section 2.3. }
  \label{Fig:abundance}
\end{figure}

The abundance gradient is an essential ingredient in an
accurate picture of galaxy formation and evolution. The existence of
abundance gradient along the Milky Way has been proved by
observations during the past twenty years using different tracers.
An oxygen or/and iron abundance gradients about
$-0.06\sim-0.07\rm\,dex\,kpc^{-1}$
were obtained by using various tracers, such as H{\sc ii} regions,
B stars (see Rudolph et al. 2006 and references therein), Planetary
Nebulae (PNe) (Maciel et al. 2006) and open clusters (Chen et al. 2003,
2008). A radial dependent SFR and an infall model with disk formed
by ``inside-out'' scenarios could well reproduce the above gradients
(Matteucci \& Francois 1989; Boissier \& Prantzos 1999; Hou et al. 2000;
Chiappini et al. 2001; Fu et al. 2009; Yin et al. 2009).

Nevertheless, the
abundance gradients for different elements are different, and the
exact gradient value, especially its time evolution, for the Milky
Way disk are still not very certain. This has prevented from a clear
constraint for the chemical evolution models. Therefore it is very
important to have more data from extragalactic galaxies. Indeed,
abundance gradients have been found in many other spiral galaxies
(Zaritsky et al. 1994). At present, the best studied extragalactic
galaxies for abundance is the M101 (Kennicutt et al. 2003), M31 and
M33 (Rosolowsky \& Simon 2008; Magrini et al. 2007b) by observing a
large sample of H{\sc ii} regions and young stars in their disks.

The observed abundance gradient of H{\sc ii} regions in M33 have
been obtained by several authors. First quantitative spectroscopic
studies were carried out by Smith (1975), Kwitter \& Aller (1981),
and V\'{i}lchez et al. (1988). These observations, which were
limited to the brightest and largest H{\sc ii} regions, implied a
steep radial oxygen gradient about $-0.1\rm\,dex\,kpc^{-1}$. Using
the data from previous observations with derived electron
temperatures, Garnett et al. (1997) re-determined the radial oxygen
abundance and obtained an overall oxygen gradient of
$-0.11\pm0.02\rm\,dex\,kpc^{-1}$, including the central regions of
M33. Recent studies seemed to converge to a much shallower
gradients, such as Crockett et al. (2006), Magrini et al. (2007b)
(hereafter M07b), Rosolowsky \& Simon (2008) (hereafter RS08) and
Magrini et al. (2010), deriving the radial oxygen gradients as
$-0.012\pm0.011\rm\,dex\,kpc^{-1}$,
$-0.054\pm0.011\rm\,dex\,kpc^{-1}$,
$-0.027\pm0.012\rm\,dex\,kpc^{-1}$, and
$-0.044\pm0.009\rm\,dex\,kpc^{-1}$, respectively.

Using PNe spectroscopy, Magrini et al. (2004)
measured the element abundances of 11 PNe in M33. Magrini et al.
(2007a; hereafter M07a) derived an oxygen radial gradient of
$-0.11\pm0.04\rm\,dex\,kpc^{-1}$, while Magrini et al. (2009;
hereafter M09) presented a new result of
$-0.031\pm0.013\rm\,dex\,kpc^{-1}$ from 91 PNe in M33. Adding their
observed data of 16 PNe to the 32 (out of 91) sample of M09,
Bresolin et al. (2010) obtained an oxygen radial gradient of
$-0.013\pm0.016\rm\,dex\,kpc^{-1}$ from this combined sample.

Abundance gradients in M33 disk have also been estimated from B-type
giants, for example, Monteverde et al. (1997) derived the oxygen
radial gradient of $-0.16\pm0.06\rm\,dex\,kpc^{-1}$, and Urbaneja et
al. (2005) obtained the oxygen radial gradient of
$-0.06\pm0.02\rm\,dex\,kpc^{-1}$.

Clearly, the true situation of the abundance gradient in M33 disk is
still not fixed. Optical line spectroscopy are mainly concentrated
on the B stars, H{\sc ii} regions. The main uncertainty comes from the
empirical calibrations which are used to derive the electron
temperatures in the nebular. And the stellar data suffers from lack
of enough sample. Also the derived abundance gradient depends on the
distance ranges. In any case, a larger and homogeneous sample of H{\sc ii}
regions which spread the whole M33 disk is needed in order to have
conclusive results about the real gradients. Such project is
currently undergoing by a couple of groups (Rosolowsky \& Simon
2008; Magrini et al. 2007b).

In Table 2, we present most currently available oxygen abundance
gradient measurements for M33 disk. We plot the oxygen abundance
observed in PNe, H{\sc ii} regions and B-type super-giants along
the disk in Fig. \ref{Fig:abundance}. The observed data derived
through PNe are denoted by open circles, where the 32 (out of 91)
red open circles are taken from M09 and the 16 green ones are
taken from Bresolin et al. (2010). The open squares represent
the observed data obtained though H{\sc ii} regions, where
the blue, the cyan, the green, the yellow, the magenta and
the red open squares are taken from RS08, M07b,
V\'{\i}chez et al. (1988), Kwitter \& Aller (1981), Crockett et al.
(2006) and Bresolin et al. (2010), respectively. In addition, the
observed data estimated from B-type super-giants are shown in the
same panel via black empty triangles. In order to compare model
predictions with the observations, these data are also plotted
in the left-bottom panel of Fig. \ref{Fig:diftau}.

A clear property from Fig. \ref{Fig:abundance} is that the M33 disk
has a sub-solar overall metallicity. It is consistent with the fact
that low mass galaxies have lower metallicity than the high mass
system, following the mass-metallicity relation among galaxies
(Tremonti et al. 2004 ). Physically, the low metal content in low
mass galaxies can be interpreted as the role of either the large gas
accretion (infall) or the strong galactic winds (outflow), because
both can change the metallicity of the galaxy disk and gas fraction
(Dalcanton 2007). Since M33 is a low mass system, we have reasons to
assume that M33 is undergoing substantial galactic outflow processes
during its evolution.

\section{The Model}
\label{sect:analysis}

Motivated by various observed properties available along the M33
disk, we build a bridge between the observations of M33 and its SFH
by constructing a chemical evolution model. We assume that the disk
has been embedded in a dark matter halo. Primordial gas
($X=0.7571, Y_{\rm p}=0.2429, Z=0$) within the dark halo cools down
gradually to form a rotationally supported disk. The disk is
basically assumed to be sheet-like and composed by a series of
independent rings with the width of 500 pc. Since radial flows
are still in the stage of lacking a well understood
description and will bring additional free
parameters and uncertainties in the model, we do not consider
radial flows in this paper. The details and essentials of
our model are as follows.

\subsection{Gas infall rate}

The gas infall process has been introduced in the formation model of
the Milky Way disk due to the well-known ``G-dwarf problem", which
means that the simple ``closed-box" model cannot explain the locally
observed metallicity distribution function of long-lived stars
(Pagel 1989). Recent observations, which were detected at 21-cm by
Westmeier et al. (2005), also showed that the M33 disk is still in
the process of accreting substantial gas.

We assume that the M33 disk is progressively built up by the infall
of primordial gas from its dark matter halo. For the given radius $r$,
the gas infall rate $f_{\rm{in}}(r,t)$ (in units of
$\rm{M_{\odot}}\,{pc}^{-2}\,{Gyr}^{-1}$) is assumed to be:
\begin{equation}
f_{\rm{in}}(r,t)=A(r)\cdot t\cdot e^{-t/\tau},
\label{eq:infall rate}
\end{equation}
where $\tau$ is the infall time-scale and it is a free parameter
in our model. The $A(r)$ are actually a set
of separate quantities constrained by the stellar mass surface density
at the present time. In practice, we iteratively estimate $A(r)$ 
by requiring the model resulted stellar mass surface
density at the present time
is equal to its observed value (Chang et al. 2010, 2012).
The stellar disk of M33 is described by an
exponential surface density profile and given by
\begin{equation}
\Sigma_*(r,t_{\rm g})=\Sigma_*(0,t_{\rm g}) {\rm exp}{(-r/r_{\rm d})},
\label{eq:surface density}
\end{equation}
where $\Sigma_*(r,t_{\rm g})$ is the stellar mass surface density at
the present time. $t_{\rm g}$ is the cosmic age and we set
$t_{\rm g}=13.5\rm\,Gyr$ according to the standard
cosmology. $\Sigma_*(0,t_{\rm g})$ is the
central stellar mass surface density at the present time.
$r_{\rm d}$ is the radial scale-length of the stellar disk and
we set $r_{\rm d}=1.4\,\rm{kpc}$.
The total stellar mass of the M33 disk is given by
\begin{equation}
M_{*}(t_{g})=\Sigma_*(0,t_{\rm g}) 2\pi r_{\rm d}^2.
\label{eq:central surface density}
\end{equation}
We adopt $M_{*}(t_{\rm g})=4.0\times10^{9}\,{\rm M}_{\odot}$
and accordingly we set $\Sigma_*(0,t_{\rm g})=325 \rm{M_{\odot}}\,{pc}^{-2}\,$.

In previous models of the chemical evolution of disk galaxies, the gas
infall rate is widely assumed to be exponentially decreasing with
time (Hou et al. 2000; Chiappini et al. 2001; Yin et al. 2009).
However, because of small initial mass of M33, the disk may initially accumulate
a small amount of gas and its accretion rate may gradually increase
as its gravitational potential builds up, and then start decreasing when the gas
reservoir is depleted (Prantzos \& Silk 1998). Therefore, we adopt another
form of gas infall rate in this paper. Generally speaking, the gas infall rate
we adopted is low at the beginning and gradually increases with time. It
reaches to the maximum value when $t=\tau$ and then slowly
falls down. We emphasize that the infall time-scale $\tau$ is an important free
parameter in our model, which regulates the shape of gas accretion
history and then largely influences the main properties of SFH along
the disk.

\subsection{Star formation law }

It is well known that almost all stars form in molecular clouds,
therefore, it is natural to expect that the SFR correlates more
strongly with the molecular gas surface density than with the total
one. Indeed, studies of spacial resolved SF law show that the SFR
correlates stronger with the surface density of molecular hydrogen
rather than that of atomic hydrogen (Wong \& Blitz 2002; Bigiel et
al.2008; Leroy et al. 2008). In this paper, we adopt the
$\Sigma_{\rm H_2}$-based SF law (see equ. \ref{eq:h2sfr}).
We calculate the molecular hydrogen depletion
time of M33 using the observed surface density of SFR and $\rm
H_2$ in Section 2.1 and we adopt $t_{\rm dep}=0.43\,\rm{Gyr}$
throughout this work.

Regarding the ratio of molecular to atomic gas surface density
of a galaxy disk $R_{\rm mol}$, Blitz \& Rosolowsky (2006) and
Leroy et al. (2008) proposed that the mid-plane pressure of
the interstellar medium (ISM) $P_{\rm h}$ alone could
determine $R_{\rm mol}(r,t)$:
\begin{equation}
R_{\rm{mol}}(r,t)={\Sigma_{{{\rm{H}}_2}}}(r,t)/{\Sigma_{{\rm{H{\sc I}}}}(r,t)}=
{\left[ {P_{\rm h}\left( r,t\right)/{P_0}} \right]^{\alpha_P} },
\label{eq:BRh2}
\end{equation}
where $P_{0}$ and $\alpha_{\rm P}$ are constants derived from
the observations. We adopt
$P_{0}=4.3\times10^{4}\,\rm{cm}^{-3}\,\rm{K}$ and $\alpha_{\rm
P}=0.92$ (Blitz \& Rosolowsky 2006).

According to Elmegreen (1989, 1993), the mid-plane pressure of the
ISM in the disk galaxies can be expressed as:
\begin{equation}
P_{\rm h}\left( r,t \right)=\frac{\pi }{2}G{\Sigma _{{\rm{gas}}}}\left( r,t \right)\left[ {{\Sigma _{{\rm{gas}}}}\left( r,t \right)
+ \frac{c_{\rm gas}}{c_{\rm star}}{\Sigma _*}\left( r,t \right)} \right],
\label{eq:elmegreenpressure}
\end{equation}
where $G$ is the gravitational constant, $c_{\rm{gas}}$ and
$c_{\rm{star}}$ are the (vertical) velocity dispersions of gas and
stars, respectively. Observations suggest that $c_{\rm{gas}}$ is a
constant along the disk and we adopt
$c_{\rm{gas}}=11\rm\,km\,s^{-1}$ (Ostriker et al. 2010), but
$c_{\rm{star}}=\sqrt{\pi Gz_{0}\Sigma _*(r)}$, where $z_{0}$ is the
scale-height of the disk and we adopt $z_{0}=0.5\,\rm{kpc}$ (Heyer
et al. 2004).

\subsection{Gas outflow rate}

As we have already mentioned in the previous section, the outflow
process may influence the evolution of M33. Indeed, Garnett (2002)
suggested that spiral galaxies with $V_{\rm
rot}\leq125\rm\,km\,s^{-1}$ may lose some amount of gas in
supernova-driven winds. Spitoni et al. (2010) also concluded that the outflow may
play a significant role especially for galaxies with the stellar
mass less than $10^{10}\,{\rm M}_{\odot}$. To explain the
observed correlation between the galactic gas-phase metallicity
and its stellar mass, Tremonti et al. (2004) and Chang et
al. (2010) suggested that it is necessary that the gas outflow
efficiency of a galaxy may increase as the galactic stellar mass decreases.
Since M33 is a low mass system with a rotation speed
about $V_{\rm rot}\approx110\rm\,km\,s^{-1}$, the gas outflow
process should paly an important role in its disk evolution.

Following Chang et al. (2010), we also assume that the outflow gas
has the same metallicity as that of ISM at that time and will not
fall again to the disk. The gas outflow rate $f_{\rm out}(r,t)$ (in
units of $\rm{M_{\odot}}\,{pc}^{-2}\,{Gyr}^{-1}$) is assumed to be
proportional to the SFR surface density $\Psi(r,t)$ :
\begin{equation}
  f_{\rm out}(r,t)=b_{\rm out}\Psi(r,t)
\label{eq:outflow}
\end{equation}
where $b_{\rm out}$ is another free parameter in our model.

\subsection{Stellar evolution and chemical evolution equations}

The updated stellar population synthesis (SPS) model of Bruzual \&
Charlot (2003), i.e., CB07, is adopted in our work, with the stellar
initial mass function (IMF) being taken from Chabrier (2003). The
lower and upper mass limits are adopted to be $0.1\,{\rm M}_{\odot}$
and $100\,{\rm M}_{\odot}$, respectively.

Regarding the chemical evolution of the disk of M33, both the
instantaneous-recycling approximation and the instantaneous mixing
of the gas with ejecta are assumed, i.e., the gas in a fixed ring is
characterized by a unique composition at each epoch of time. We take
the classical set of equations of galactic chemical evolution from
Tinsley (1980):
\begin{equation}
\frac{d[\Sigma_{\rm tot}(r,t)]}{dt}=f_{\rm{in}}(r,t)-f_{\rm{out}}(r,t),
\end{equation}
\begin{equation}
\frac{d[\Sigma_{\rm gas}(r,t)]}{dt}=-(1-R)\Psi(r,t)+f_{\rm{in}}(r,t)-f_{\rm{out}}(r,t),
\end{equation}
\begin{eqnarray}
\frac{d[Z(r,t)\Sigma_{\rm gas}(r,t)]}{dt}=y(1-R)\Psi(r,t)-Z(r,t)(1-R)\Psi(r,t) \nonumber\\
+Z_{\rm{in}}f_{\rm{in}}(r,t)-Z_{\rm{out}}(r,t)f_{\rm{out}}(r,t),
\end{eqnarray}
where $\Sigma_{\rm tot}(r,t)$ is the total (star + gas) mass surface density.
$Z(r,t)$ is the metallicity in the ring centered at
galactocentric distance $r$ at evolution time $t$. $R$ is the return
fraction and we set $R=0.3$
according to the adopted IMF. $y$ is the stellar yield and we set
$y=1\,{\rm Z}_{\odot}$ (Chang et al. 2010). $Z_{\rm{in}}$ is the metallicity
of the infalling gas and we assume the infalling gas is primordial,
that is $Z_{\rm{in}}=0$. $Z_{\rm{out}}(r,t)$ is the metallicity
of the outflowing gas and we assume that the outflow gas
has the same metallicity as that of ISM, e.g.,
$Z_{\rm{out}}(r,t)=Z(r,t)$ (Chang et al. 2010).

   \begin{figure*}
   \centering
   \includegraphics[angle=0,height=14cm,width=0.75\textwidth]{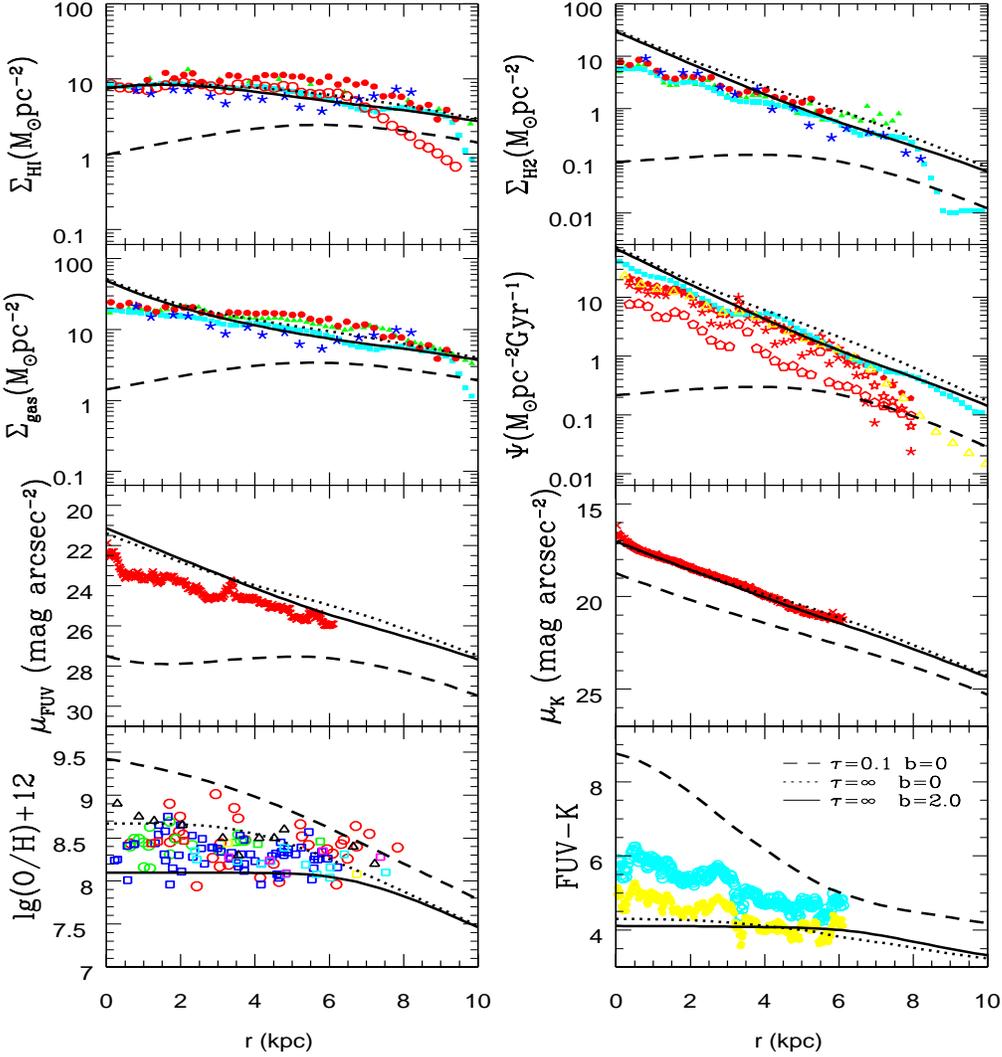}
   \caption{Influence of free parameters on the model predictions. At the left-side,
    from top to bottom it shows the H{\sc i} surface density, total gas surface density,
    surface brightness in FUV$-$band, and oxygen abundance radial profiles.
    At the right-side, from top to bottom it shows H$_{2}$ surface density,
    SFR surface density, surface brightness in $K-$band, and FUV$-K$ color
    radial profiles. Different line types correspond to various parameter
   groups: dashed lines $(\tau, b_{\rm out})=(0.1\,{\rm Gyr}, 0)$, dotted lines
   $(\tau, b_{\rm out})=(\infty$, 0), solid lines $(\tau, b_{\rm out})=(\infty$, 2).
   The observed data are described in details in Section 2. }
   \label{Fig:diftau}
   \end{figure*}

In summary, two free parameters in our model are the infall
time-scale $\tau$ and the outflow coefficient $b_{\rm out}$. The
combination of gas infall rate and outflow rate determines the
behavior the total (gas+star) mass surface density. We assume that
the initial total mass surface density is zero, then we can
numerically obtain the total mass surface density at any time after
free parameters being given. Adding the SF law, we can easily describe
how much cold gas turns into stellar mass and then numerically calculate the
chemical and color evolution of the disk of M33.

\section{Model Results Versus Observations}


   \begin{figure*}
   \centering
   \includegraphics[angle=0,height=14cm,width=0.75\textwidth]{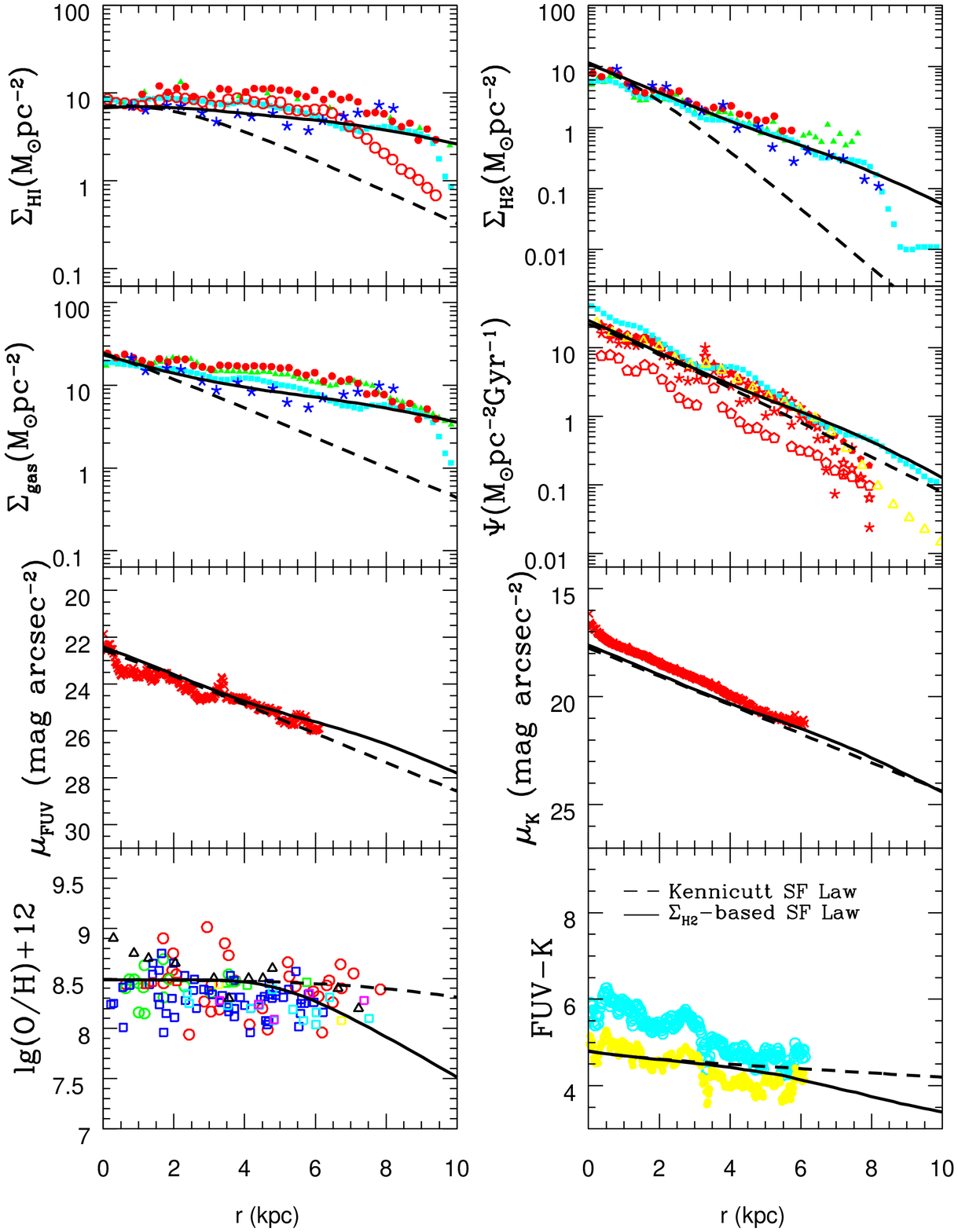}
   \caption{Comparison of the predictions of the viable model with
   the observations. The solid lines plot the viable model results,
   which adopts $(\tau, b_{\rm out})=(r/{\rm r_d}+5.0\,\rm{Gyr}, 0.5)$
   and with the $\Sigma_{\rm H_2}$-based SF law, while the dashed
   lines are with the same $(\tau, b_{\rm out})$ but with the Kennicutt
   SF Law $\Psi(r,t)=0.25\Sigma_{\rm gas}^{1.4}(r,t)$. The
   observed data symbols are the same as those in Fig. \ref{Fig:diftau}. }
   \label{Fig:difSFR}
   \end{figure*}

Figure. \ref{Fig:diftau} presents comparison between the model
predictions and the observations. The left-side of Fig.
\ref{Fig:diftau} shows the H{\sc i} surface density, total gas
surface density, surface brightness in FUV$-$band, and oxygen
abundance radial profiles. The right-side shows H$_{2}$ surface
density, SFR surface density, surface brightness in $K-$band, and
FUV$-K$ color radial profiles, respectively. The data of observed radial
profiles of H{\sc i} are taken from Corbelli (2003) (green filled
triangles), Heyer et al. (2004) (cyan filled squares) and
Gratier et al. (2010) (blue asterisks). The data taken from
Verley et al. (2009) are shown by red empty circles
(Westerbork) and filled circles (Arecibo). The left-second panel of
Fig. \ref{Fig:diftau} plots the total gas surface density, which
is defined as $1.33(\Sigma_{\rm{H_{2}}}(r,t)+\Sigma_{\rm HI}(r,t))$ (the
factor of 1.33 considers the contribution of helium), where the data
notation is the same as that of H$_2$. The data of total gas surface
density taken from Verley et al. (2009) is calculated using the Arecibo data
of $\Sigma_{\rm HI}$ except for the first four inner radius,  where we used the
values from Westerbork. The notations of other panels of the observed data are
the same as that described in details in Section 2.

We explore the influence of free parameters on model results
step by step. Firstly, we do not consider the contribution of
gas outflow process ($b_{\rm{out}}=0$), and explore the
influence of $\tau$ on model results. The dashed and dotted lines in
Fig. \ref{Fig:diftau} denote the model predictions of two limiting cases of
$\tau=0.1$\,Gyr and $\tau\rightarrow\infty$, respectively. The case of
$\tau\rightarrow0$ corresponds to a time-declining infall rate,
which is very close to the ``closed-box" model, while
$\tau\rightarrow\infty$ corresponds to a nearly constant with a slight
time-increasing gas infall rate.

It can be seen from Fig. \ref{Fig:diftau} that the model predictions
are very sensitive to the adopted
$\tau$. The model adopting a shorter $\tau$ predicts lower gas surface
density (i.e. both H{\sc i} and H$_{2}$ components), lower SFR,
lower surface brightness, redder color, and higher metallicity than
that of adopting longer $\tau$. This is mainly due to the fact that,
in our model, the setting of short infall time-scale corresponds to a
fast gas infall process and high SF process in the early time of the
galaxy evolution, and then leads to old stellar population, high
metallicity and low cold gas content at the present day.

To investigate the influence of gas outflow process on the evolution of M33,
the solid lines in Fig. \ref{Fig:diftau} plot the model results
adopting $\tau=\infty$ and $b_{\rm{out}}=2$. Comparison between
solid lines and dotted lines shows that the gas outflow process has
no significant influence on the stellar population and the cold gas
content, but it does reduce the gas-phase metallicity since it takes
a fraction of metals away from the disk. Therefore, the observed
radial distribution of oxygen abundance may tightly
constrain the gas outflow process of M33.

Another interesting point of Fig. \ref{Fig:diftau} is that the area
between the dashed and solid lines almost covers the whole region of
the observations, which means that it is possible to construct an
evolution model that can reproduce the main features of the
observations of M33 disk. Considering the observed trends that the
inner stellar disk seems to be redder and metal richer than that of
the outer region, we adopt the inside-out disk formation scenario,
i.e., a radial dependent infall time-scale $\tau=r/{\rm
r_d}+5.0\,\rm{Gyr}$ and a moderate outflow rate $b_{\rm
out}=0.5$ as the viable model, and its
results are plotted as solid lines in Fig. \ref{Fig:difSFR}. The
observed data is the same as that of the Fig. \ref{Fig:diftau}. It
can be seen that the solid lines are in fairy agreement with the
most of observational data, which suggests that our viable model
includes and reasonably describes the important ingredients
of main processes that regulate the formation and evolution of the M33.

The infall time-scale is one of the important free parameters
in our model. After exploring its influences on the radial profiles
of M33 and computing several sets of combinations of free
parameters, we finally choose the radial-dependent infall time-scale in
the viable based on the balance that the model predictions can be
consistent with most of the observed data. We emphasize that although the accurate values of
free parameters in the viable model are not unique, the main trend that
$\tau(r)$ increases from the inner disk to the outer parts are robust. In fact,
this assumption is consistent with the well-known idea that the disk forms inside-out
and has already applied in previous models of formation and evolution of disk
galaxies (Matteucci \& Francois 1989; Boissier \& Prantzos 1999; Hou et al. 2000;
Chiappini et al. 2001; Fu et al. 2009; Yin et al. 2009).

  \begin{figure}
   \centering
   \includegraphics[angle=0,height=6cm,width=8cm]{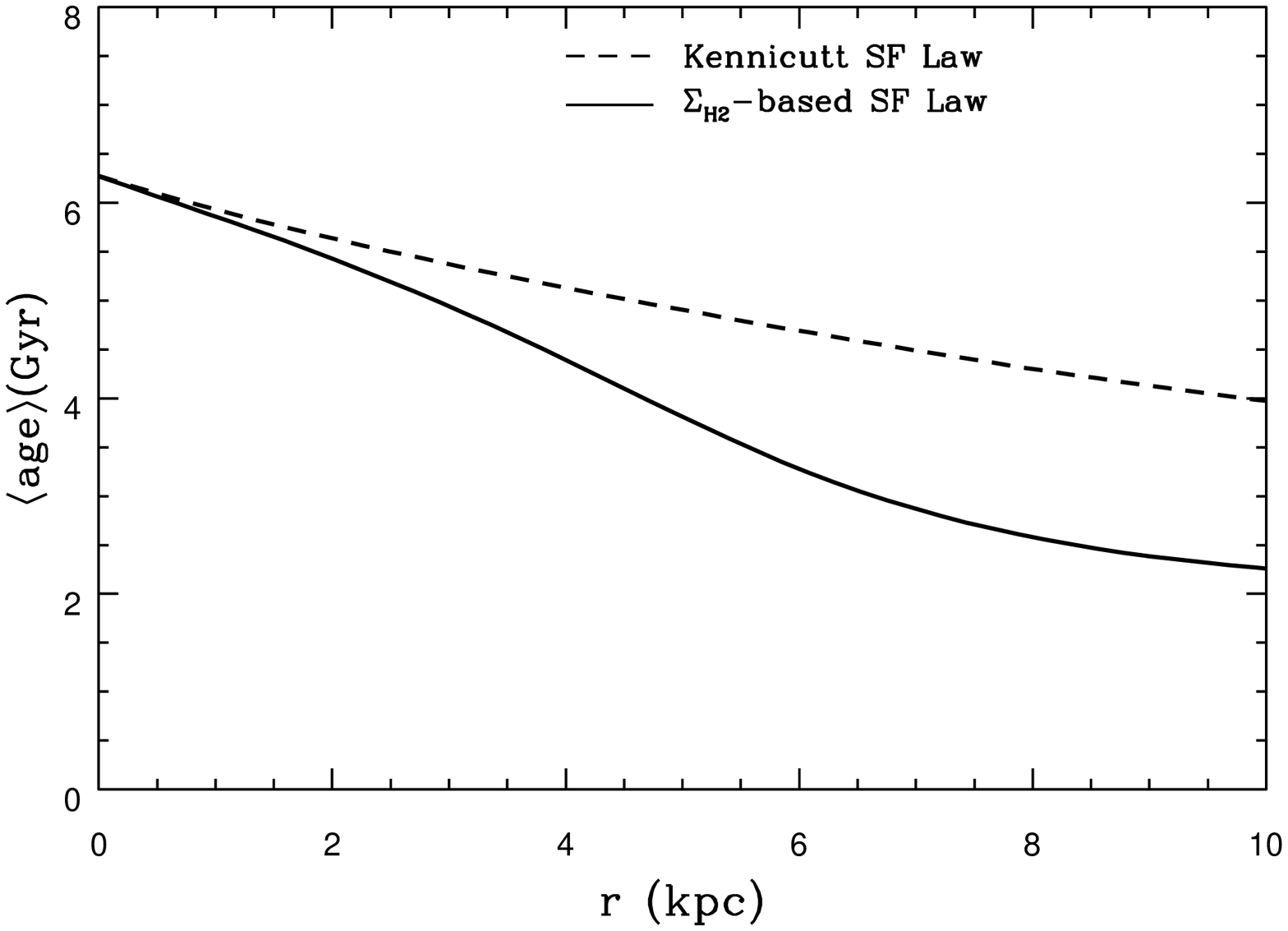}
   \caption{Current radial profiles of mean stellar age with different
   SF laws, solid line ($\Sigma_{\rm H_2}$-based SF law), dashed line
   (Kennicutt SF law).}
   \label{Fig:age}
   \end{figure}

In order to clearly demonstrate the property
of the inside-out formation scenario, we plot the mean stellar age
along the disk predicted by our typical model with solid lines in
Fig. \ref{Fig:age}. It can be seen that the model predicts a
decreasing mean stellar age from the inner disk to the outer region
since the short infall time-scale means a large fraction of stars
formed at early stage and hence old mean stellar age. Indeed, the
ground-based observations of bright stars in M33 suggest that the
outer disk of M33 may have formed at late epoch and the SF
process may be on-going (Davidge 2003; Block et al. 2004;
Rowe et al. 2005).

Another method to test the inside-out formation scenario is to
measure the radial distribution of the mean age of the stellar
populations along the disk. Williams et al. (2009) presented
resolved stellar photometry of four fields along the M33 disk and
found that the age of the majority of the stars decreases with
increasing galactic distance, which is consistent with our model
prediction. On the other hand, the investigation of stellar age
distribution in far outer disk of M33 (outside the break radius
$r>8\rm kpc$) shows that the mean age of stars in far outer disk is
young ($\sim 3\rm Gyr$) but the age increases with radius (Barker et al.
2007; Barker et al. 2011). The explanation of the origin of this
reverse age gradient in the far outer disk of M33 is still an open
question and needs further investigations.

The SF law is another important components of the model that may
largely influence the evolution of M33. In this paper, we adopt
the $\Sigma_{\rm H_2}$-based SF law instead of the classic
Kennicutt SF law (that is, $\Psi(r,t)=0.25\Sigma_{\rm gas}^{1.4}(r,t)$)
in previous studies. We compare the model predictions when using these
two kinds of SF laws in Fig. \ref{Fig:difSFR} and Fig. \ref{Fig:age}.
The results of Kennicutt SF law are shown by dashed lines.
Other ingredients of the model, including the
molecular-to-atomic ratio $R_{\rm mol}$, are the same as
that of the viable model. It can be seen that the solid lines are
clearly distinguished from the dashed lines, especially in the outer disk
where the cold gas surface density is low. Comparing to the model
adopting the $\Sigma_{\rm H_2}$-based SF law, the model adopting the
Kennicutt SF law predicts much flatter gradients of metallicity,
mean stellar age and color, while steeper gradients of $\Sigma_{\rm
H_2}$, $\Sigma_{\rm HI}$ and $\Psi(r,t)$. These results
suggest that, comparing to the $\Sigma_{\rm H_2}$-based SF law, the
SFE of Kennicutt SF law is higher in the outer regions of the disk.
This means that the cold gas in the outer regions will turn
into stars more efficiently, which will result in an older stellar
population. Indeed, previous works on the evolution of the Milky Way
and M31 disks have shown that, in order to agree with the observed
metallicity gradient and its time evolution well, it is necessary to
adopt a modified Kennicutt SF law, such as $\Psi(r,t) \propto
\Sigma_{\rm gas}^{1.4}(r,t)/r$ (Boissier \& Prantzos 1999; Chang et al
1999; Fu et al. 2009; Yin et al. 2009). However, why the SFE should
be inversely correlated directly with the galactic radius is not
fully understood. Our results show that the model predictions based on
the $\Sigma_{\rm H_2}$-based SF law are more consistent with
observed trends, especially the radial distributions of both the
cold gas and the stellar population.

\section{Summary}
\label{sect:summary}

In this paper, we construct a parameterized model of the
formation and evolution of M33 disk by assuming that the disk
originates and grows by the primordial gas infall. The gas infall
rate is described by a simple formula with one free parameter, the
infall time-scale $\tau$. We also include the contribution of gas
outflow. A molecular hydrogen correlated SF law is adopted to
describe how much the cold gas turns into the stellar mass. We
numerically calculate the evolution of M33 and compare the model
predictions with the observational data.

The main results of our model can be summarized as follows:

\begin{enumerate}

\item Based on the observed $\Sigma_{\rm H_2}$ and
the SFR, we estimate the depletion time of molecular
hydrogen $t_{\rm dep}$ along the disk of M33. It is shown
$t_{\rm dep}$ does not vary very much with the radius, which
suggests that the SFE of M33 is almost constant along the disk.
We also show that the SFE of M33 is higher than the
average value derived by Leroy et al. (2008) on the basis of a large
sample of galaxies.


\item Our results show that the model predictions are very
sensitive to the adopted infall time-scale. A long infall
time-scale will result in blue colors, low metallicity, high
H$_{2}$ and H{\sc i} mass surface densities, high SFR surface
density.

\item We also find that the outflow has relatively little
effects on the disk stellar population and cold gas content. But it
has great influence in shaping the abundance profiles along the M33
disk since it takes a fraction of metals away from the M33 disk
due to its low mass potential.

\item The model which adopts a moderate outflow rate and an inside-out
formation scenario, that is, the infall time-scale increases with radius,
can be in good agreement with the most of observed constraints of M33.
Our results suggest that the formation of M33 is quiet and it may not
form through violent accretion process.

\item It is shown that the model adopting the Kennicutt SF law predicts
much flatter gradients of color, metallicity and mean stellar age
and steeper gradients of cold gas than that adopting the
$\Sigma_{\rm H_2}$-based SF law. Our results imply that, comparing
to the Kennicutt SF law, the $\Sigma_{\rm H_2}$-based SF law would be
more suitable to describe the evolution of the galactic disk,
especially for the radial distributions of both the cold gas and the
stellar population.

\end{enumerate}

\section*{acknowledgements}
We thank the referee for the suggestions to greatly improve this work.
We thank Simon Verley, Pierre Gratier, Fabio Bresolin and
Juan Carlos Mu\~{n}oz-Mateos for kindly providing some observational
data of M33. This work was funded by the National Natural Science
Foundation of China under No. 11103058, 11173044, 11033008,
10821061, the Key Project No. 10833005 and No. 10878003, the Group
Innovation Project No. 11121062, the Science Foundation of Shanghai No.
11ZR1443600, and the Knowledge Innovation Program of the Chinese
Academy of Sciences.

{}

\bsp
\label{lastpage}
\end{document}